# Bottom Stresses of Static Packing of Granular Chains

*Guan Wang[1] Degan Hao[1] Ning Zheng[1] [*] Liangsheng Li[2][†] Qingfan Shi[1][**]*

1) School of Physics, Beijing Institute of Technology, Beijing 100081, China

2) Science and Technology on Electromagnetic Scattering Laboratory, Beijing 100854, China

We experimentally measure the static stress at the bottom of a granular chains column with a precise and reproducible method. The relation, between the filling mass and the apparent mass converted from the bottom stress, is investigated by using various chain lengths. Our measurements reconfirm that the scaling behavior of the stress saturation curves is in accord with the theoretical expectation of the Janssen model. Additionally, the saturation mass is displayed as a nonmonotonic function of the chain length, where a distinguishing transition of the saturation mass is found at the persistence length of the granular chain. In order to understand the transition of saturation mass, the friction coefficient and the volume fraction of granular chains are also measured, from which Janssen parameter can be calculated.

[*] ningzheng@bit.edu.cn
[†] liliangsheng@gmail.com
[**] qfshi123@bit.edu.cn

I. INTRODUCTION

The static packing of granular materials has been one of long-lasting issues due to its practical importance in many fields including civil engineering, soil science, and storage and processing of raw materials, while the unity of basic physics remains unsolved [1]. Among the studies attempting to address the puzzle, generally most researches deal with spherical grains [2-4], and occasionally the grains with aspherical shapes such as ellipsoids [5, 6] or rods [7, 8] have also been used to alter the packing structure. Whatever grains are used in investigating the packing properties, nearly all researches have focused on individual grains. Until recently granular chains began to draw more attention in the researches on the packing. For instance, Zou et al. investigated the static packing structure of granular chains in a cylinder, finding the reciprocal of the volume fraction of chains packing is analogous to the glass transition temperature of polymer solutions. The similar behavior between jamming state of granular chains with glass state of polymers was thus illustrated [9]. Our experiments subsequently revealed that statistical scaling characteristics of two-dimensional granular chains are in good accord with the theoretical expectations of polymer models [10-12].

The previous contributions on the static packing of granular chains mainly put emphasis on the packing structures near the jamming transition [9, 13], or on an analogy between their statistical properties and theoretical predictions in polymer science [9-12], aiming to provide a possible way to uncover the physics that is experimentally inaccessible for polymers. Although the packing column confined in a container can also act as a simple but valid experimental system to test many theoretical models for the repartition or the transmission of stresses in granular materials [14-18], the mechanical behavior of the packing of granular chains did not receive enough attention until Brown et al revealed that a stress response in granular chains was very distinct from common grains, namely individual grains [19]. They found that strain stiffening occurred in the packing of long chains due to the inter-chains entanglements, which are absent in the packing system of individual grains or short chains. However, the lateral wall of the silo in their experiment is a soft membrane that can allow for free radial expansion when applying a compressive stress. The investigation on the mechanical properties of confined granular chains with a *rigid*, *lateral boundary* appears to be still scarce. Measuring the stresses under various boundary conditions is very necessary for the establishment of the macroscopic constitutive relations of stress-strain.

In this paper, we accurately measure the average mass at the bottom of a granular chain column confined in a rigid cylinder, with the measurement method proposed by Vanel et al [20]. In the measurement, eleven different chains, consisting of the same material but the chain length $N$ spanning three orders of magnitude from $N$=2 to 2048, are used to measure the average mass at the bottom of the chains column as a function of the granular chains filling mass, respectively. The relation between them is in a good agreement with the description of the Janssen model. We thus apply the Janssen model to fit the experimental curves, and then extract the saturation masses for all chains. The saturation mass as a nonmonotonic function of the chain length is also shown, where a transition of the saturation mass is found at the persistence length of the granular chain. Moreover, these relevant parameters to determine the saturation mass are measured, and the Janssen parameter is calculated for all chains as well. Finally, qualitative arguments for the nonmonotonic dependence of the saturation mass are presented.

**II. Experimental setup**

A diagrammatic sketch of the experimental establishment is illustrated in Fig. 1. The semi-rigid granular chains are confined in an acrylic cylinder with an inner diameter of 45 mm and a height of 700 mm. The cylindrical silo is vertically mounted to a heavy stand, and its bottom is closed by a movable piston connected to a force sensor which records the average pressure on the piston. The diameter of the piston is slightly smaller than the inner diameter of the silo in order that the piston never touches the inner wall, and the gap between the piston and the wall is small enough to avoid the leakage or the blockage of filled materials. The linear chains with free ends used in our experiment are composed of hollow, steel beads of a diameter $3.0\pm0.1$ mm and links with the same material. The link between two neighboring beads can stretch out and draw back freely, the maximum length of which is about $1.2 \pm 0.1$ mm. The chain length $N$ is represented by the number of the beads on a chain. The number of the beads required to form a smallest ring, which is generally equivalent to the persistence length of the chain, $\xi = 8$ is used to characterize the stiffness of the granular chain, as the inset in Fig. 1 shows. Due to hollow structure, the measured apparent density of the granular chain is not equal to its solid density, but $3.56 \pm 0.12 g/cm^3$.

A filling mass $M_{fill}$ of granular chains is poured into the silo through a hopper. We also balled a long chain into the silo, finding little influence on the bottom stress measurement provided the rest of the measuring procedure remains identical except for filling methods. After the filling procedure is completed, the stepping motor controls the piston to move downward at a constant speed $V = 0.2 mm/s$. The piston would not stop to wait 90 seconds until it goes through a 2 mm downward displacement. The downward motion allows the granular column to slide down so that the friction along the inner wall is supposed to be fully mobilized. During the waiting time, we make use of a measuring protocol proposed by Vanel et al, which can precisely and reproducibly obtain the mean pressure at the bottom of a granular column [20]. Then, the piston continues its downward movement and the bottom stress measurement repeats. When a total displacement 20 mm ends, a complete stress measurement is fulfilled. After the complete measurement, the silo is entirely emptied and refilled. Hence, data from every trial correspond to completely independent sets of the experiment.

Following a complete stress measurement, a thin and flat cardboard is used to flatten the top surface of the chains column carefully and then a well-defined surface is obtained. The volume fraction of the chains column is estimated by monitoring its height $H$. The volume fraction $\nu$ is given by

$$\nu = \frac{M}{H\pi R^2} / \rho_a \qquad (1)$$

Where $M$ is the filling mass, $H$ is the height of the chains column, $R$ is the radius of the silo, and $\rho_a$ is the apparent density measured. To eliminate other undesirable influences which disturb the measurement as much as possible, temperature and humidity are maintained at $23\pm1^oC$ and 45%-50%, respectively.

### III. Results and discussion

In the bottom stress measurement, the chain length ranges three orders of magnitude from $N=2$ to 2048. According to the persistence length $\xi=8$, these chains may be categorized into two groups, short chains ($N \leq \xi$) and long chains ($N > \xi$). Fig.2 (a) and (b) show the relationship between the apparent mass, which is converted from the bottom stress, and the filling mass for short and long chains, respectively. In Fig.2, all curves approach their own saturation values. For short chains, the saturation mass increases with the chain length. However, the situation is reverse for long chains.

For the asymptotic saturation of the filling mass in a silo, Janssen proposed a pioneering work to explain the counterintuitive phenomenon [21]. The model treated a granular packing column as a continuous medium, and assumed that the friction between the grains and the wall was fully mobilized along the upward direction, that thus the part of the vertical stress is transferred to the horizontal orientation. The apparent mass $M_a$ at the bottom of the silo as a function of $M_{fill}$ is in the form of the equation as following

$$M_a = M_{sat}[1 - \exp(\frac{M_{fill}}{M_{sat}}) \quad \text{with} \quad M_{sat} = \frac{\rho \nu \pi R^3}{2K\mu_s} \quad (2)$$

Where $\rho$ is the density of granular materials, $\mu_s$ is the friction coefficient between the silo wall and the granular materials, $K$ is the Janssen parameter, redirecting the part of vertical stresses to horizontal direction via the friction, $R$ is the radius of the silo and $\nu$ is the volume fraction of the granular column. To verify that the experimental curves can be described by the Janssen model, we rescale the apparent mass and the filling mass with the saturation mass. It is found that although the experimental curves separate in Fig.2 (a) and (b), all rescaled data cluster together to form a universal scaling curve which agrees with the one predicted by Janssen model: $M_a/M_{sat} = f(M_{fill}/M_{sat})$, with $f(x) = 1 - \exp(-x)$, see Fig.2 (c).

In Fig.2, the saturation masses for different chains appear to change with the chain length. Thus, the saturation mass is plotted as a function of the chain length, shown in Fig.3. The relation between the saturation mass with the chain length is a nonmonotonic function, showing a peak at the persistence length. Within the range of short chains, the saturation mass increases with the chain length, finally reaching a maximum at $N = \xi$. But at the range of long chains, the saturation mass is a monotonically decreasing function of the chain length.

In the Janssen model, the saturation mass can be expressed by $M_{sat} = \dfrac{\rho v \pi R^3}{2K\mu_s}$. To understand the nonmonotonic behavior of the saturation mass, the chain length dependences of the packing volume $v$ and the friction coefficient $\mu_s$ are measured. It is clearly observed that the volume fraction $v$, which reflects the average structure properties of the packing column, declines monotonically with the chain length $N$, from $v \approx 0.61$ for $N = 2$ to a saturation value $v \approx 0.42$ at large $N$ on the characteristic curve, as shown in Fig.4 (a). The curve in Fig.3 exhibits a nonmonotonic characteristic, but the chain length dependence of the volume fraction is a monotonic function. Consequently, even if the saturation mass depends on the volume fraction, it is not the only parameter to determine the nonmonotonic behavior. From the expression of the saturation mass, the saturation mass also relates to the friction coefficient $\mu_s$ and the Janssen parameter $K$, besides the volume fraction $v$. We measure the friction coefficients of different chains using the sliding angle of a chains sledge. Fig.4 (b) shows that the friction coefficient is almost independent on the chain length, basically a constant. From the relationship of the saturation mass, the friction coefficient and the volume fraction, the Janssen parameter $K$ can be calculated. It is found in Fig.4 (c) that in short chains case $K$ reduces with the chain length $N$, but in long chains case the situation is opposite. A dip appears at the persistence length as well. However, Janssen parameter *K* is only an effective parameter derived from the experimental data, which cannot be acquired by direct measurements. Thus the underlying physics of the nonmonotonic relation still remains ambiguous.

Recently strain stiffening was observed in the granular chains column as the chains packings were sheared in uniaxial compression [19]. Specifically, the distinct mechanical responses are divided by the persistence length. While the chain length was shorter than the persistence length, all measuring results exhibited typical granular strain softening. As the chain length was longer than the persistence length, the chains column showed strong strain stiffening that further enhanced with the increase of the chain length. The reason that strain stiffening only occurred for the longer chains system could be attributed to the inter-chains entanglements.

Compared with Brown et al results [19], we also find a distinguishing transition of the saturation mass at the persistence length. For the packings of the long chains ($N>8$), the inter-chains entanglements become increasingly common and significantly influence the packing structure, leading to change of the mechanical response of the packing system. In contrast, the short chains ($N \leq 8$) preferentially behave more like individual flexible rods with different aspect ratios, due to the absence of inter-chains entanglements. Even if the chain $N=8$ can be bent into a smallest ring ideally, it has a very small probability to form a loop in the packings, as observed in the experiment. Correspondingly, it is plausible to suspect the inter-chain entanglements might be responsible for the transition of the saturation mass. The findings of ours and Brown et al imply that the chains packings solely with different chain lengths are supposed to be seen as the packings of various materials, at least on some mechanical aspects, although all mechanical properties of these chains such as density and friction coefficient are identical. However, the questions whether the inter-chains entanglements are indeed responsible for the transition of the saturation mass, and how to explain the underlying physics, remain open for future work.

## IV. Summary

In conclusion, we fill a rigid silo with granular chains, and measure the average stress at the bottom by using a precise and reproducible method. The apparent mass, that is, measured mass at the bottom increases with the filling mass until reaching a plateau noted saturation mass. We use eleven kinds of granular chains with different chain lengths to test the relation between the apparent mass with the filling mass. When being scaled with the saturation mass, the measured data from different chains collapse together to form a universal master curve which agrees with the prediction of the Janssen model. We employ the Janssen model to fit the experimental curves, and then the saturation masses are extracted. The chain length dependence of the saturation mass appears to be nonmonotonic. As the chain length is shorter than the persistence length ($N<8$), the saturation mass increases with the chain length. However, if the chain length is longer ($N>8$), the saturation mass is inversely proportional to the chain length. The saturation mass reaches a peak at $N=8$. The Janssen parameter $K$ is calculated by using the relationship of the saturation mass, the volume fraction and the friction coefficient. The relation between the Janssen parameter $K$ and the chain length $N$ is nonmonotonic as well. For the nonmonotonic behavior, we speculate that the short chains can be seen as anisotropic, elastic rods due to the absence of inter-chains entanglements, showing the mechanical response similar to the common grains. But for the long chains, the amount of the entanglements is proportional to the chain length, and the existence of numerous inter-chains entanglements might introduce a qualitative difference of packing structures from the short chains system. However, the question whether inter-chains entanglements are indeed responsible for the nonmonotonic behavior of the saturation mass remains open.

ACKNOWLEDGMENTS

The work was supported by the National Natural Science Foundation of China (Grants No. 11104013) and the National Innovative Experimental Projects for University Students (Grant No. 201210007044).

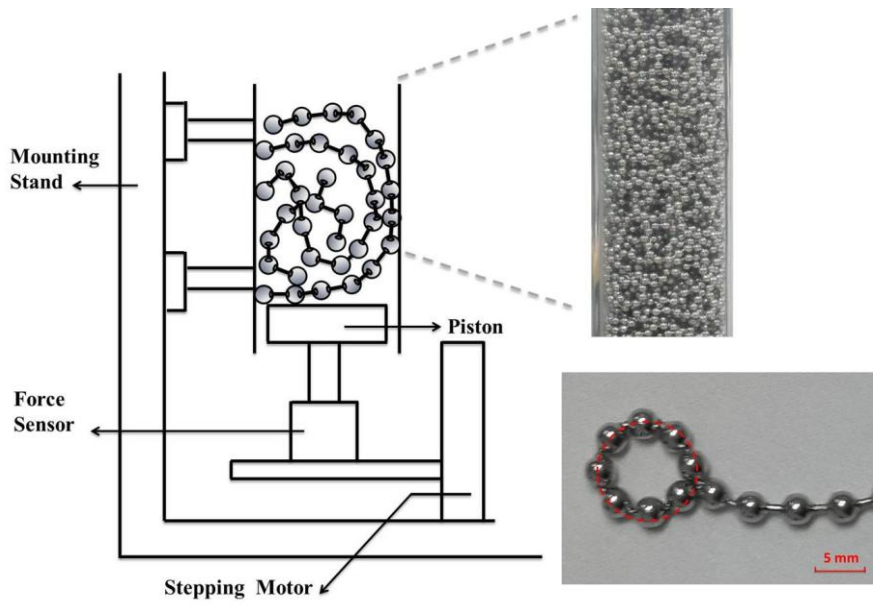

FIG.1. (Color online) Sketch of the experimental set-up (not to scale). An acrylic silo clamped vertically is filled with granular chains. The bottom of the silo is closed by a piston attached to a force sensor. The diameter of the piston is slightly smaller than that of the silo such that the design can avoid the friction between the piston and the inner wall, as well as the leakage of granular chains. The enlarged portions of the schematic show a static packing of the chains with the chain length $N=2048$ (upper right), and the persistence length, $\xi=8$, namely the number of beads required to form a minimum loop which is highlighted by a red dashed circle (lower right).

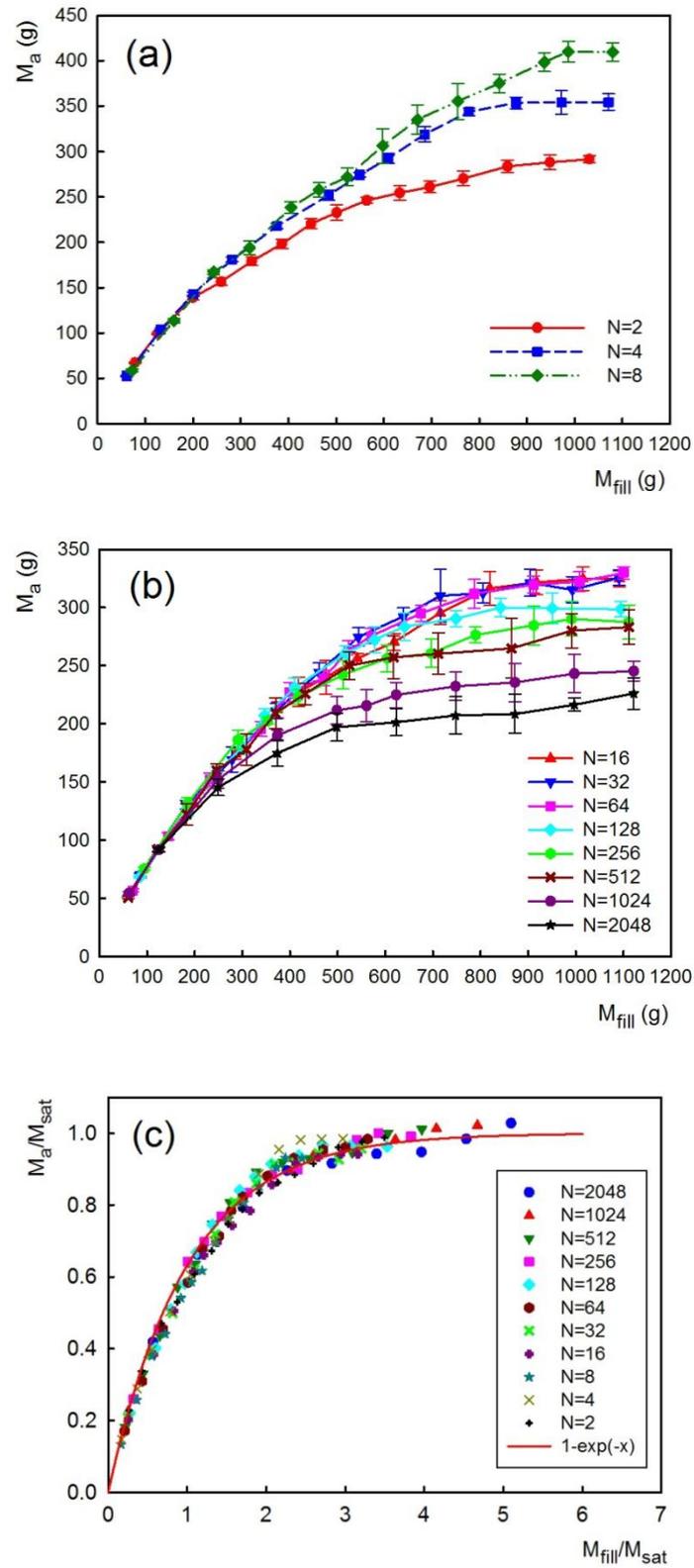

FIG.2. (Color online) (a) Apparent mass $M_a$ as a function of filling mass $M_{fill}$ is plotted for short chains with different chain lengths $N$=2 (circle), $N$=4 (square), and $N$=8 (diamond). (b) Apparent mass as a function of filling mass is plotted for long

chains with different chain lengths $N=16$ (up triangle), $N=32$ (down triangle), $N=64$ (square), $N=128$ (diamond), $N=256$ (Hexagon), $N=512$ (cross), $N=1024$ (circle), and $N=2048$ (star). All data points in (a) and (b) have been averaged about 50 times from 7 experiment trails. (c) Apparent mass $M_a$ vs filling mass $M_{fill}$, rescaled by the saturation mass $M_{sat}$, for all chain lengths. All data points cluster together to form a universal scaling curve that agrees with the prediction of the Janssen model (solid line).

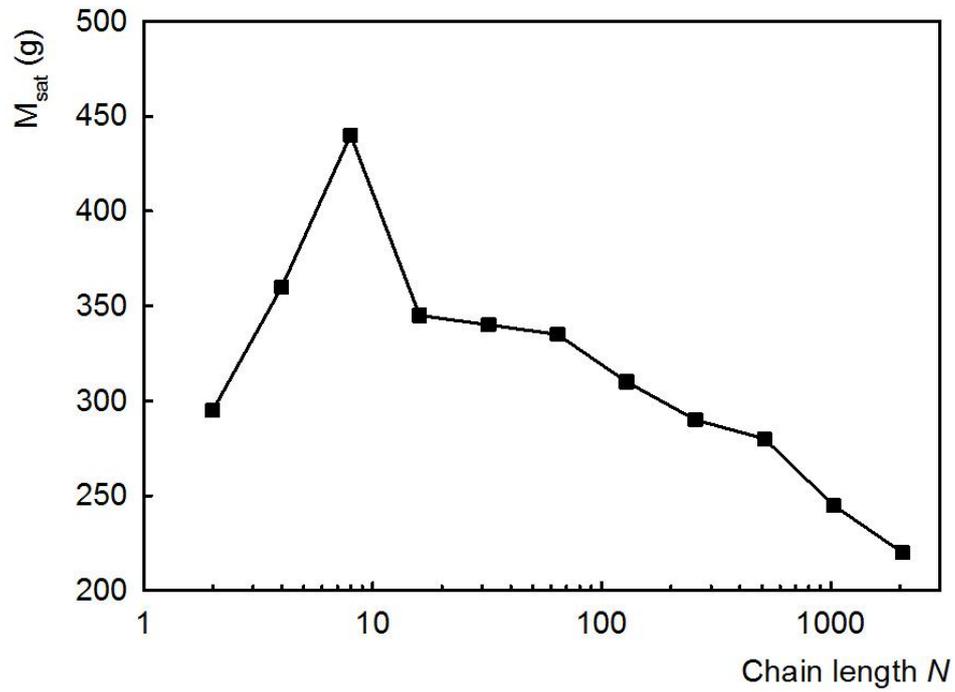

FIG.3. Chain length dependence of the saturation mass that is extracted from Janssen model is plotted using a semilogarithmic scale. The peak corresponds to the persistence length $\xi = 8$ at *x* axis. The solid line is plotted for eye guide.

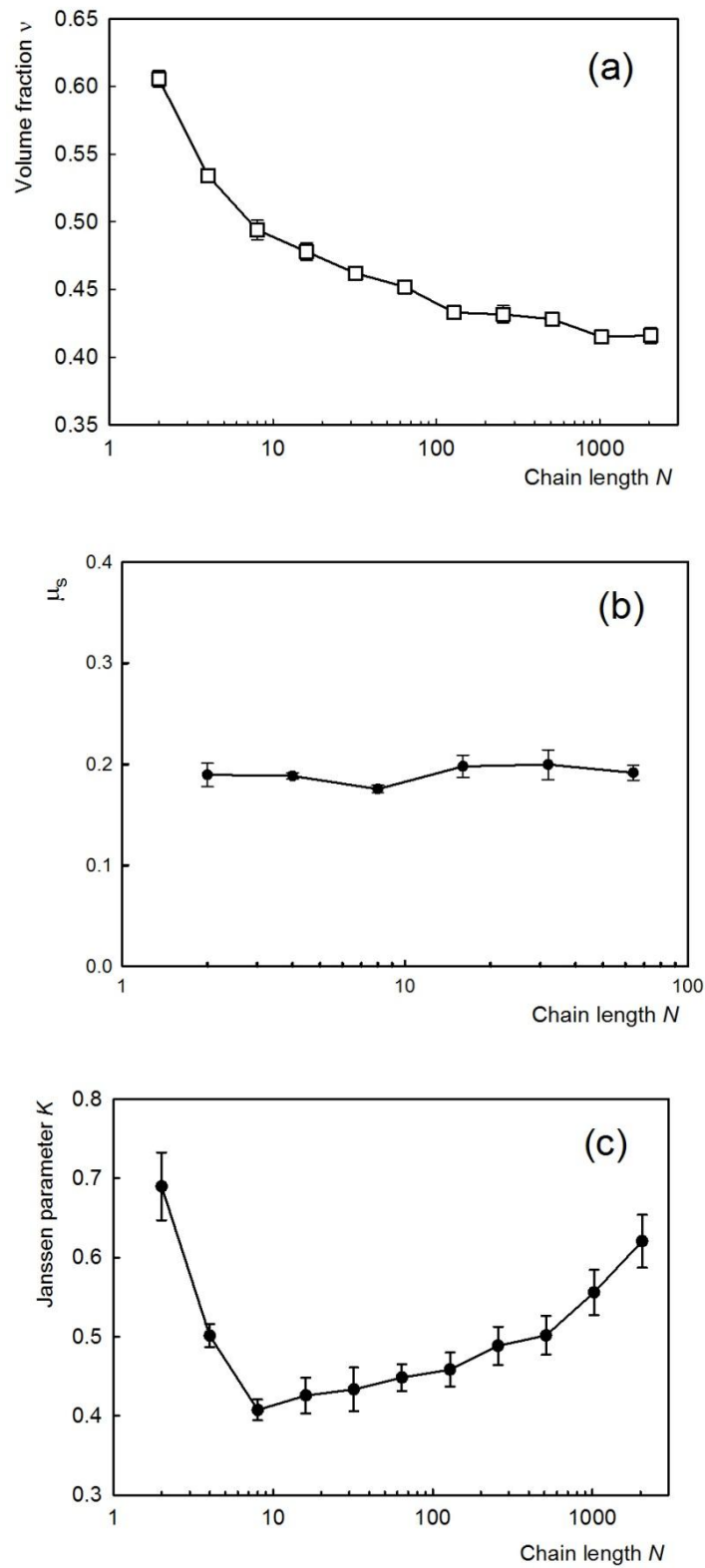

FIG.4. (a) Volume fraction $v$ Vs the chain length $N$, using a semilogarithmic scale. (b) Friction coefficient $\mu_s$, between the granular chains and the inner wall of the silo, as a function of the chain length $N$ is shown using a semilogarithmic scale. (c) Janssen

parameter $K$ as a function of the chain length $N$. The minimum on the curve corresponds to the persistence length $\xi = 8$.